\documentclass[conference]{IEEEtran}
\IEEEoverridecommandlockouts
\usepackage{cite}
\usepackage{amsmath,amssymb,amsfonts}
\usepackage{algorithmic}
\usepackage{graphicx}
\usepackage{textcomp}
\usepackage{xcolor}
\usepackage{soul}
\def\BibTeX{{\rm B\kern-.05em{\sc i\kern-.025em b}\kern-.08em
    T\kern-.1667em\lower.7ex\hbox{E}\kern-.125emX}}
\begin{document}

\title{XWAVE: A Novel Software-Defined Everything Approach for the Manufacturing Industry
\thanks{This work has been supported by the XWAVE research grant (KK-2025/00056), funded by the Basque Government.}
}

\makeatletter
\newcommand{\linebreakand}{%
  \end{@IEEEauthorhalign}
  \hfill\mbox{}\par
  \mbox{}\hfill\begin{@IEEEauthorhalign}
}
\makeatother

\author{
\IEEEauthorblockN{Juanjo Zulaika, Ibone Oleaga}
\IEEEauthorblockA{\textit{Tecnalia} \\
San Sebastian, Spain \\
juanjo.zulaika,ibone.oleaga@tecnalia.com}
\and
\IEEEauthorblockN{Anne Sanz}
\IEEEauthorblockA{\textit{Ideko} \\
Elgoibar, Spain \\
asanz@ideko.es}
\and
\IEEEauthorblockN{Naia Presno}
\IEEEauthorblockA{\textit{Tekniker} \\
Eibar, Spain \\
naia.presno@tekniker.es}
\linebreakand
\IEEEauthorblockN{Aitor Landa-Arrue, Miguel Barón}
\IEEEauthorblockA{\textit{Ikerlan Technology Research Centre} \\
Arrasate/Mondragón, Spain \\
alanda,mbaron@ikerlan.es}
\and
\IEEEauthorblockN{María del Puy Carretero}
\IEEEauthorblockA{\textit{Vicomtech} \\
San Sebastian, Spain \\
mcarretero@vicomtech.org}
\and
\IEEEauthorblockN{Unai Lopez-Novoa}
\IEEEauthorblockA{\textit{University of the Basque Country} \\
Bilbao, Spain \\
unai.lopez@ehu.eus}
}

\maketitle

\begin{abstract}
The manufacturing sector is moving from rigid, hardware-dependent systems toward flexible, software-driven environments. This transformation is shaped by the convergence of several Software-Defined technologies: Software-Defined Automation virtualizes industrial control, replacing proprietary PLCs with containerized, programmable solutions that enable scalability and interoperability. Software-Defined Compute and Communications provide a means to distribute intelligence seamlessly across devices, networks, and cloud platforms, reducing latency and enabling dynamic reconfiguration. Software-Defined Manufacturing Systems, usually implemented as Digital Twins, are real-time virtual models of machines and processes, allowing predictive analysis, optimization, and closer integration between human operators and intelligent systems. This work presents XWAVE, a project that unites these three Software-Defined paradigms to present a modular, fully software-defined manufacturing system. 
\end{abstract}


\section{Introduction}

The manufacturing industry operates with inherently rigid architectures in which Planning, Automation, and Communication operations are executed on physical equipment with dedicated connections and software. This means that whenever a company wishes to scale, adapt, or reconfigure an element or component of its Manufacturing Systems, it must physically modify those systems. This reality imposes unavoidable limitations on the manufacturing sector in terms of customization, efficiency, sustainability, resilience, and productivity \cite{ADEL2024101431}.

The global industry is on the threshold of a new industrial revolution, driven by the convergence of advanced automation, the virtualization of production systems, ubiquitous and integrated connectivity. In this context, the Software-Defined Everything (SDx) paradigm proposes the virtualization of key functionalities of different resources and assets, thereby enabling their scalability, configuration, reconfiguration, and customization\cite{10609236}. 

Based on these recent SDx paradigms, this work presents XWAVE, a project whose goal is to accelerate this global trend toward the virtualization and software definition of diverse industrial equipment. The project is executed by a consortium of 5 research centres (Tecnalia, Ideko, Tekniker, Ikerlan and Vicomtech) and the University of the Basque Country, which proposes to conceive, within a two-year timeframe, a first conceptual laboratory-scale environment in which all industrial equipment dedicated to Automation, Communications, and the Connection of Manufacturing Systems is fully virtualized and managed through software.

XWAVE aims to achieve this goal by addressing three complementary SDx technologies: Software-Defined Automation (SDA), Software-Defined Communications and Computing (SDC$^3$), and Software-Defined Manufacturing Systems (SDMSs). While each of these areas has evolved independently, XWAVE seeks to integrate them into a unified framework that leverages virtualization, programmability, and interoperability. The objective is to provide a holistic solution that enhances flexibility, scalability, and resilience in manufacturing environments.

This paper presents an analysis of the state of the art in each of the three domains, highlighting both the opportunities and the limitations that currently hinder their industrial adoption. For each area, the paper outlines the vision of the XWAVE project, detailing how the proposed approaches can overcome existing shortcomings and contribute to next-generation manufacturing systems. 

The remainder of this paper is structured as follows: Section~\ref{sec:areas} presents an analysis of the state of the art of SDA, SDC$^3$ and SDMSs. Based on the previous analysis, Section~\ref{sec:xwave} describes the XWAVE approach for each domain, and finally, Section~\ref{sec:conc} presents some final conclusions.

\section{Software-Defined Everything} \label{sec:areas}

This section presents the different Software-Defined technologies that XWAVE will be based on, including a brief analysis of the state of the art and identified gaps and limitations.

\subsection{Software-Defined Automation}

SDA is described as the integration of open-source technologies and standards into industrial processes to control multiple hardware assets from containerized software running on hardware resources~\cite{LIN2025114076}. One of the most innovative trends is the concept of the Virtual PLC, whose main idea is to emulate the traditional functions of a PLC outside of its dedicated hardware, enabling execution on PCs or electronic boards~\cite{Gaffurini_2024}.

The virtualization of PLC applications will facilitate their deployment in production lines through the use of containers, thereby accelerating commissioning~\cite{Koziolek_2024}. In this regard, Docker has become the most widely used deployment tool in recently published virtual PLC developments~\cite{Francia_2024} and, currently, its use is being analysed in industrial systems requiring real-time operations \cite{Diogenes_2022}. At the connectivity level, these works employ communication protocols such as Zenoh, OPC UA and Ethernet/IP to ensure interoperability and integration across different devices.

A revelant use case of virtual PLCs is provided by Susen~\textit{et al.}~\cite{Susen_2024}, who present the concept of a virtualized numerical control operating in the cloud. The focus is on synchronizing such numerical controllers with field devices by employing EtherCAT and commercial hardware for computing and networking. These results support the concept of deterministic cloud-based numerical control powered by open-source software.

While SDA promises flexibility, efficiency, and scalability in industrial automation, it faces key challenges such as interoperability, latency, security, and the need for deterministic response:

\begin{itemize}

\item Insufficient Integration of SDA with Emerging Technologies: The SDA concept still struggles to fully leverage Artificial Intelligence (AI), Digital Twins (DTs), advanced robot programming software (ROS2), and Edge Computing to optimize industrial processes.

\item Low Reliability and Resilience in Industrial Environments: Unlike industrial PLCs, which are designed for demanding environments, SDA software-based solutions are not yet capable of meeting the much higher robustness requirements in connectivity.

\item Insufficient Integration of SDA with Industrial IoT (IIoT) Systems: Soft-PLCs are not yet able to efficiently integrate with IIoT platforms and ensure compatibility with containerization tools such as Docker or industrial communication protocols like OPC UA and MQTT.

\item Presence of Cybersecurity Gaps: Since Soft-PLCs are software-based and network-connected, they are more vulnerable to cyberattacks, compounded by weak security-by-design and early-lifecycle processes to prevent future code vulnerabilities.

\item Insufficient Real-Time Performance: Neither high workload environments nor non-deterministic operating systems such as Windows and Linux currently guarantee precise, deterministic, real-time control —a critical requirement for future industrial processes to be managed by Virtual-PLCs.

\end{itemize}

\subsection{Software-Defined Communication and Computation Continuum}

SDC$^3$ is a concept that stems from the ``Edge-to-Cloud Continuum''~\cite{Milojicic_2020} paradigm, also known as the ``Computing Continuum''~\cite{AlDulaimy_2024}. Conceived as a mechanism that enables flexible distribution of data acquisition and intelligence, the key idea is to abstract the concrete implementation of the communication infrastructure in a particular environment. In industry, where Cyber-Physical Systems (CPS) demand ultra-low latency networks as well as extreme reliability and high adaptability, the ability to define and manage connectivity and computing infrastructures in a programmable manner is fundamental for the transition toward a Full Software-Defined Factory Network~\cite{Koyasako_2023}.

At a low level, SDC$^3$ setups rely on technologies such as Software-Defined Networking (SDN), P4 (a domain-specific programming language for network devices), and virtualized edge applications orchestrated by systems with control-plane access. New lightweight application-layer protocols, such as Zenoh~\cite{Baron_2025}, transport-layer technologies, such as QUIC~\cite{Fan_2023}, and deterministic networking mechanisms like DetNet~\cite{Ahmed_2024}, are emerging in the industrial communications domain as enablers of reconfigurable, robust, and low-latency communications.

Network virtualization, particularly in 5G networks through the concept of network slicing, further separates infrastructure from services, enabling multiple configurations and modules over the same physical hardware~\cite{Patel_2022}. Unlike traditional networks, virtualization allows multiple flow treatments within a single node, reusing links for different services with distinct requirements. 

With this evolution, a new challenge arises: independent resource control. The use of consolidated industrial protocols alongside new solutions such as Zenoh and Fast DDS enables the creation of dynamic data flows, optimizing communication in heterogeneous environments~\cite{Scanzio_2021} and ensuring interoperability. 

With all this, the use of SDC$^3$ infrastructures in industry faces multiple challenges, among which the following have been identified as most relevant:

\begin{itemize}
\item Lack of interoperability between platforms and closed standards: This limitation reduces application portability in fully distributed environments beyond Edge vs Cloud computing.

\item Insufficient flexibility in resource management and communication performance optimization: There are currently no industrial applications, often dependent on proprietary technologies, that allow the integration of communication resource management systems in open and highly dynamic ecosystems.

\item Insufficient orchestration of networks and computing: Current network orchestration capabilities are insufficient to guarantee strict Quality of Service (Hard-QoS) in dynamic manufacturing environments.

\item Absence of standardized mechanisms for intelligent roaming systems in industry: There are currently no solutions in IIoT environments that automatically and transparently select the most suitable network at each moment based on the location and requirements of the connected device.

\item Insufficient security in distributed infrastructures: There are currently no advanced solutions for distributed infrastructures, such as AI-based Software-Defined Firewalls, nor dynamic defense strategies such as Moving Target Defense (MTD).

\end{itemize}

\subsection{Software-Defined Manufacturing Systems} 

SDMSs have usually been tightly coupled with DTs, in which a system or physical process can be synthesized as a virtual representation that receives real-time data from its physical counterpart, enabling simulation, analysis, and optimization in virtual environments~\cite{Gaffinet_2025}. 

A successful application of DTs usually requires robust and scalable solutions to maximize their potential. In this regard, the use of microservices and micro-frontends enables full connectivity across different DTs as well as the development of customized presentation layers~\cite{Simoes_2024}. This approach helps decompose complex and distributed systems, manage supplier diversity, and facilitate service orchestration, enabling real-time data exchange between the virtual model and physical machinery.

The implementation of DTs in Manufacturing Systems presents different maturity levels within the current state of the art. At the base are system-of-systems models, where machines communicate with other machines in real time, data is interoperable, and standards of classification, ontology, and semantic relationships are adopted. This enables predictive operations, real-time analytics, simulation, and other augmented operations~\cite{Watt_2024}. Achieving this requires that data be integrated using a common data model, such as the Unified Name Space (UNS) reference framework.

In recent years, several studies have addressed the process of creating DTs for Manufacturing Systems. Some approaches are IT-based, such as applications, app stores, and application execution systems~\cite{Miclaus_2016}, while others are based on the composition and integration of DTs as systems-of-systems~\cite{Michael_2022}. In this context, Hasan~\textit{et al.}~\cite{Hasan_2020} propose the creation of blockchain-based DTs for Manufacturing Systems, offering several interfaces such as RestHTTP, Web3, or JSON RPC.

The MakeTwin reference architecture~\cite{Fei_2024} is presented as a universal software platform for DT creation. Meanwhile, \cite{LinX_2023}~study the impact of modular systems in manufacturing, which can facilitate the transfer of software developed by manufacturers to suppliers. This approach allows manufacturers to access supplier applications such as downloadable components or cloud services and integrate them into their internal systems, providing greater flexibility, customization, and the potential for a pay-per-use model. In parallel, manufacturers can efficiently leverage external expertise and resources, gaining access to specialized solutions while maintaining cost-effectiveness. Along these lines, \cite{Simoes_2024}~proposes an approach based on micro-frontends and microservices, while \cite{Yang_2025}~present an approach relying solely on microservices.

Identified shortcomings in the state of the art of SDMSs are:

\begin{itemize}
\item Insufficient Realism of DTs: Current 3D models are not visually comparable to the real entity, nor do the virtual entities behave equivalently to the physical one. As a result, user perception is degraded because personal experiences cannot be correlated with the virtualization of the process.

\item Insufficient Interoperability among DTs: At present, it is not possible to interconnect DTs from different manufacturers or companies. This leads to inefficient redundancy in systems with multiple providers, generating dependency on those providers.

\item Insufficient Interoperability among Digital Twin Deployment Platforms: Currently, there is no solution that ensures automated and automatable maintenance of the entire SDMS, since each DT is deployed on its own platform.

\item Insufficient Cybersecurity in DTs: At present, no solution guarantees full security of communications between physical manufacturing systems and their DTs, or among the DTs themselves. Security is delegated to the communication channel, resulting in a loss of control over the data read, managed, and generated by the DTs.
\end{itemize}

\section{The XWAVE approach} \label{sec:xwave}

Building upon the analysis of the current state of the art, XWAVE defines a set of technological priorities and research directions aimed at overcoming the identified shortcomings.

In the domain of SDA, XWAVE proposes to address the identified shortcomings by:

\begin{itemize}
\item Integrate and deploy software-based PLCs in Dockerized environments to ensure control of advanced functionalities in a manufacturing asset.

\item Design and implement a modular and open architecture for advanced manufacturing automation solutions that facilitates scalability.

\item Conceive and design a software-based PLC control algorithm with deterministic latency.

\item Integrate open communication protocols such as Zenoh to guarantee system interoperability with minimal latency.

\item Conceive the integration of software-based PLCs with Docker on an Edge device and on general-purpose hardware.

\item Integrate security practices into the Continuous Integration/Continuous Deployment(CI/CD) pipeline, incorporating automated tools for code analysis, penetration testing, and security audits at every stage of development.

\end{itemize}

In regards to SDC$^3$, the XWAVE project proposes to:

\begin{itemize}
\item Conceive interoperability solutions for different platforms and standards in the industrial domain, enabling data availability in distributed application environments.

\item Develop advanced orchestration mechanisms that guarantee strict QoS in dynamic industrial environments.

\item Design an intelligent system that calculates, at each moment, the optimal balance between latency, reliability, and flexibility in SDN with 5G and future 6G network architectures.

\item Design and develop Federated Learning algorithms for detecting deviations in network or device behavior.

\item Create cyber deception environments for attacker isolation, ensuring a rapid and coordinated response to mitigate the impact of any communication incident.

\item Optimize industrial communications by designing solutions based on new lightweight protocols such as Zenoh or QUIC, multipath techniques, and deterministic technologies like DetNet or Time Sensitive Networking (TSN).

\item Develop control loops based on monitoring and telemetry that allow real-time network configuration adjustments according to demand and system status.
\end{itemize}

Finally, in the domain of SDMS, XWAVE proposes to:

\begin{itemize}
\item Increase the realism of DTs by automating the conversion of engineering CAD models into interactive, photorealistic 3D models in Universal Scene Description (USD) format.

\item Advance in the standardization challenges of DTs by abstracting them into ``black boxes'' that present a known interface: inputs, outputs, requirements, and behavior.

\item Conceive an intelligent system to adapt internal networks to provide the necessary bandwidth. Develop an intelligent mechanism to identify the elements that consume the most bandwidth and propose simplification processes to reduce usage.

\item Design algorithms that minimize computation times for DTs without compromising accuracy.

\item Automate the deployment and commissioning of DTs through standards-based architectures and suitable internal networks.

\item Delegate cybersecurity to the intrinsic security of the communication channel, concealing the circulation of production data over the network.
\end{itemize}

The vision of the project is that, by 2027, the XWAVE consortium will present a first conceptual demonstrator of a scalable and modular Manufacturing System, fully software-based and capable of being dynamically reconfigured through software, without the need for any physical modification.

\section{Conclusions} \label{sec:conc}

The convergence of SDA, SDC$^3$, and SDMSs is redefining the future of industrial manufacturing. The state of the art highlights both the transformative potential and the persistent shortcomings in terms of interoperability, real-time performance, resilience, and cybersecurity. With all this as a foundation, XWAVE seeks to accelerate the transition toward flexible, scalable, and secure factories.


\bibliographystyle{IEEEtran}
\bibliography{biblio}

\end{document}